\documentclass[aps,prl,twocolumn,groupedaddress,showpacs,preprintnumbers,amsmath,amssymb]{revtex4}

\usepackage{graphicx}
\usepackage{amsmath}
\usepackage{amsfonts}
\usepackage{color}

\begin{document}

\title{Mesoscopic Fluctuations of Coulomb Drag of Composite Fermions}

\author{A. S. Price$^1$, A. K. Savchenko$^1$, D. A. Ritchie$^2$}
\affiliation{$^1$School of Physics, University of Exeter, Exeter EX4 4QL, UK} \affiliation{$^2$Cavendish Laboratory,
University of Cambridge, Madingley Road, Cambridge CB3 0HE, UK}

\begin{abstract}
We present the first experimental study of mesoscopic fluctuations of Coulomb drag in a system with two layers of composite fermions, which are seen when either the magnetic field or carrier concentration are varied. These fluctuations cause an alternating sign of the average drag. We study these fluctuations at different temperatures to establish the dominant dephasing mechanism of composite fermions.
\end{abstract}

\maketitle

Coulomb drag is a powerful technique for measuring directly the strength of electron-electron (\textit{e-e}) interaction. Coulomb drag studies are performed on two closely spaced but electrically isolated layers, where a current is driven through one of the layers (active layer) and the voltage drop is measured along the other (passive) layer. The origin of this voltage is a momentum transfer between the charge carriers in the two layers via \emph{e-e} interaction. Coulomb drag is well studied theoretically \cite{JauhoPRB93,ZhengPRB93,KamenevPRB95,FlensbergPRB95,LevchenkoPRL08} and to date has been used in a range of experimental systems \cite{GramilaPRL91,SivanPRL92,GramilaPRB93,HillPRL97,RubelPRL97,LillyPRL98,PillarisettyPRB05,YamamotoScience06,PriceScience07}.

The majority of the Coulomb drag experiments have been performed on electron or hole systems in the absence of a magnetic field. A specific case in Coulomb drag occurs in the presence of strong magnetic fields when all electrons are contained within the lowest Landau level and the filling factor $\nu$ has a value of $1/2$. Under these conditions the strongly interacting electrons can be described by composite-fermion quasiparticles, each of which representing an electron coupled with two magnetic flux quanta $\Phi_0 = h/e$ \cite{StormerRevModPhys99}. In previous theoretical \cite{UssishkinPRB97} and experimental \cite{LillyPRL98} studies, the Coulomb drag in composite-fermion systems has been shown to be strongly enhanced and have a different temperature dependence compared with the case of $B=0$.

Recently, it has been shown that Coulomb drag in disordered systems can demonstrate the interplay of \emph{e-e} interactions and quantum interference \cite{NarozhnyPRL00}. This was observed as reproducible fluctuations of the drag \cite{PriceScience07}. These have a similar origin as the reproducible conductance fluctuations \cite{LeePRB87} that have long been studied in single layer systems, and arise from the interference between electrons in each layer. Unlike the conductance fluctuations, the fluctuations of the drag resistance are remarkable in that they can exceed the average drag, resulting in random changes in the sign of the drag as the carrier concentration and magnetic field are varied.

The properties of composite fermions around $\nu = 1/2$ are similar in many respects to those of normal non-interacting electrons at small $B$-field \cite{CFbook}. Indeed, there has been a theoretical prediction that a fluctuating drag is also expected in a system of composite fermions \cite{NarozhnyPRL01}. The properties of these fluctuations depend upon the dephasing mechanisms of composite fermions.

We report here the observation of the fluctuations of the Coulomb drag between composite fermions. Despite the significant increase in the magnitude of drag of composite fermions relative to that of normal electrons, the fluctuations of the drag can still exceed the average, resulting in an alternating sign of the drag.

\begin{figure}[h]
\begin{center}\leavevmode
\includegraphics[width=0.9\linewidth]{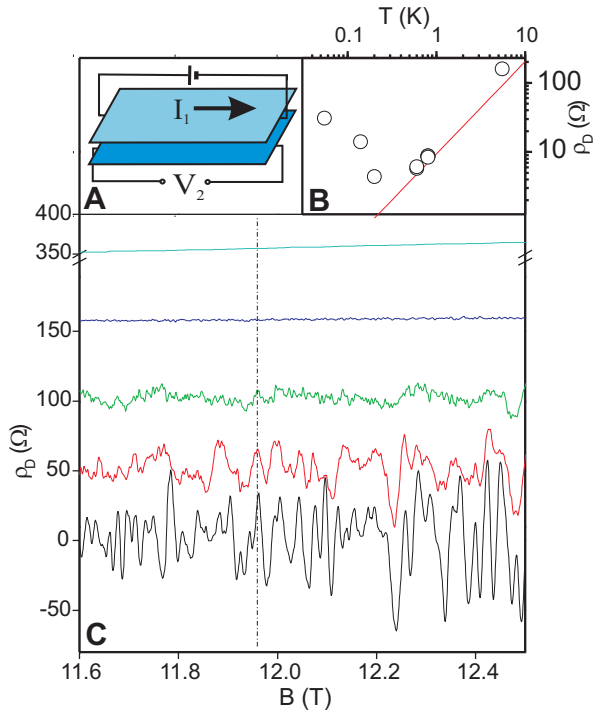}
\caption{Panel A: Schematic of the measurement circuit used to measure the Coulomb drag. Panel B: The drag resistivity as a function of temperature, at a fixed $B$-field of 11.96 T. The solid line is Eq. \ref{AvgDragEq}. Panel C: Drag resistivity as a function of magnetic field for different temperatures; $T$ = 0.05, 0.14, 0.2, 0.8, 5.6 K, from bottom to top. The graphs are offset from each other by 50 $\mathrm{m\Omega}$ for clarity. The concentration of each layer is $n = 1.45\times 10^{11}\,\mathrm{cm^{-2}}$ such that $\nu = 1/2$ at $B_{1/2} = 12$ T. The vertical line represents the points plotted in panel A.}
\label{AvgFit}
\end{center}
\end{figure}

The samples studied in this work are AlGaAs-GaAs double-layer structures \cite{LinfieldSST93}, where the carrier concentration of each layer can be independently controlled by gate voltage over the range of $n = 0.4$ -- $2.0 \times 10^{11}\,\mathrm{cm^{-2}}$, with a corresponding change in the mobility from $1.2$ -- $6.7\times 10^5 \, \mathrm{cm^2V^{-1}s^{-1}}$. The GaAs quantum wells are $200\,${\AA} in thickness, and are separated by an $\mathrm{Al_{0.33}Ga_{0.67}As}$ layer of thickness $300\,${\AA}. Each layer has a Hall-bar geometry, $60\,\mathrm{\mu m}$ in width and with a distance between the voltage probes of $60\,\mathrm{\mu m}$. The measurement circuit of the drag voltage $V_2$ is shown in Fig. \ref{AvgFit}A. The drag resistance $R_D$ is found from the ratio of the drag voltage to the current $I_1$ passed through the active layer, $R_D = -V_2/I_1$.

The drag resistance as a function of magnetic field, $\rho_D(B)$, is shown in Fig. \ref{AvgFit} for various temperatures in the vicinity of $\nu = 1/2$. A cross-section of the curves is taken at a fixed $B$-field, indicated by the dotted vertical line, and is plotted in Fig. \ref{AvgFit}B. The solid line is a plot, without adjustable parameters, of the expected value of the drag resistance of composite fermions \cite{UssishkinPRB97} in a macroscopic sample:
\begin{equation}
\rho_D = 0.825(h/e^2)(T/T_0)^{4/3}. \label{AvgDragEq}
\end{equation}
\noindent Here $T_0 \approx \pi e^2 n d / \epsilon$ = 330 K, $\epsilon$ is the dielectric constant, $n$ is the carrier concentration, and $d=500\mathrm{\AA}$ is the spacing between layers. One can see that at high temperatures, $T>$  1 K, the drag resistance is in good agreement with Eq. \ref{AvgDragEq}, and its $T$-dependence is similar to that seen in \cite{LillyPRL98} where the average drag of composite-fermions was measured. However, as $T$ is decreased the temperature dependence changes: as $T$ decreases $\rho_D$ can either decrease or increase as in Fig. \ref{AvgFit}B, depending on the carrier concentration. This non-monotonic $T$-dependence can be accounted for by the competition between the average drag and mesoscopic effects. The average drag dominates at high temperatures and decreases with decreasing temperature, whilst the amplitude of the mesoscopic fluctuations of $\rho_D$ increase with decreasing temperature, and dominate at low temperatures.

In Fig. \ref{AvgFit}C it is seen that at high temperatures the drag resistance does not contain visible fluctuations, but at lower temperatures fluctuations appear and below $T \sim 200$ mK the fluctuations dominate the drag resistivity. Note that the magnitude of these fluctuations is greatly enhanced, by a factor of $\sim 1000$, in comparison to those seen in weak $B$-fields \cite{PriceScience07}, where fluctuations were of the order of $20\,\mathrm{m \Omega}$.

\begin{figure}[h]
\begin{center}\leavevmode
\includegraphics[width=1.0\linewidth]{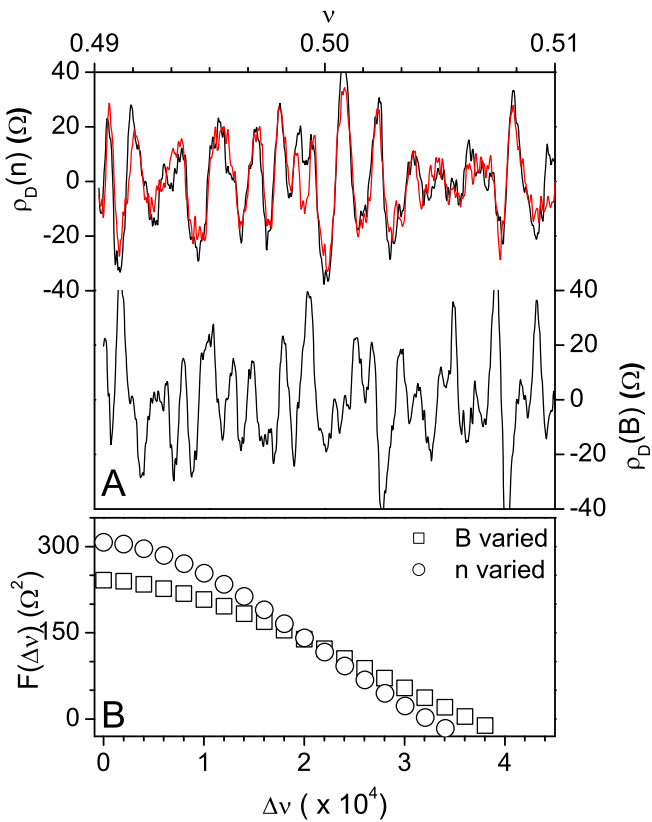}
\caption{Panel A: comparison of the fluctuations of the drag resistivity as a function of $\nu$ when $n$ and $B$ are
varied. The similarity of their periods in $\nu$ is proof that the drag is arising between interfering composite
fermions. A repeat measurement of $\rho_D(n)$ is shown to give an indication of the reproducibility of
fluctuations. $T = 50\,\mathrm{mK}$. Panel B: The autocorrelation functions of the fluctuations shown in the top panel for $\rho_D(B)$ (solid) and $\rho_D(n)$ (dashed).}
\label{ACF}
\end{center}
\end{figure}

Figure \ref{ACF} shows how the fluctuations in $\rho_D$ are also observed when the concentration of both layers is varied simultaneously, upper curve in panel A. The lower graph shows the fluctuations of the drag resistance whilst $B$ is varied and $n$ is held constant. It is clear that the fluctuations are of a similar amplitude in both experiments. It is also interesting to note that when the fluctuations are plotted as a function of filling factor $\nu$ they have a similar ``period'' $\nu_c$. This characteristic scale is determined by finding the autocorrelation function of the fluctuations, $F(\Delta \nu) = \langle \rho_D(\nu) \rho_D(\nu+\Delta \nu)\rangle$, and then taking the half-width of the half-maximum of the peak of this function. The autocorrelation functions of the fluctuations in Fig. \ref{ACF} are shown in Fig. \ref{ACF}B.

The close values of $\Delta \nu_c$ found from $\rho_D(n)$ and $\rho_D(B)$ is an important result that is expected from the flux attachment of $2 \Phi_0$ to each electron \cite{FalkoPRB94} and is a proof that the charge carriers in our system are composite fermions.

Mesoscopic fluctuations depend on both change in the Fermi energy and magnetic field. Composite fermions experience a reduced effective magnetic field that is dependent upon the external magnetic field and the density of carriers: $B^* = B - 2\Phi_0 n$. Consequently, mesoscopic fluctuations for composite fermions observed when varying the carrier density result not only from the change in the Fermi energy $\Delta E_F$, but also from the change in the effective $B$-field, $\Delta B^*(n) = 2 \Phi_0 \Delta n$ \cite{NarozhnyPRL01}. Fluctuations due to the first mechanism occur over a scale of $\Delta n_c = (\hbar/\tau_\phi)\varrho = g_{cf}/L_\phi^2$, where $\varrho$ is the density of states of composite fermions, and $g_{cf}$ is their dimensionless conductance. The second mechanism results in fluctuations on a scale of $\Delta B^*_c = 2 \Phi_0 \Delta n = \Phi_0 / L_\phi^2$. Thus, comparing the order of magnitudes of the two scales we find that fluctuations due to changes in $E_F$ occur over a scale which is $g_{cf}$ times larger than fluctuations due to the change in $B^*(n)$, so that $\Delta n_c \approx (\partial B^*/\partial n) \Delta B^*_c = 1 / 2 L_\phi^2$.

The effect of varying the external $B$-field is the same for composite fermions as it is for conventional electrons in weak $B$-fields: the correlation magnetic field $\Delta B_c$ corresponds to one magnetic flux quantum through a coherent area $L_\phi^2$. This results in the relationship between the correlation magnetic field and correlation concentration near $\nu = 1/2$: $B_c/n_c = 2 \Phi_0$. (This relation has been seen in the case of a single-layer system in which conductance fluctuations were measured near $\nu = 1/2$ \cite{KvonPRB97}).

In \cite{NarozhnyPRL01} the variance of the drag fluctuations in the ``diffusive'' regime of drag (where the mean free path is much shorter than the distance between the layers, $l \ll d$) is predicted to be
\begin{equation}
\langle \rho_D^2 \rangle \approx \frac{h^2}{e^4}\frac{1}{g_{cf}^4(\kappa d)^2}\left(\frac{L_\phi^{cf}}{L} \right)^2,
\label{eqVarDiffDrag}
\end{equation}
\noindent where $\kappa$ is the inverse Thomas-Fermi screening length, $L_\phi^{cf}$ is the coherence length of composite fermions, and $L$ is the size of the square sample. Near $\nu = 1/2$ the effective magnetic field $B^*$ is small and $g_{cf}$ is simply related to the inverse of the longitudinal resistance: $g_{cf} = (h/e^2)(R_{xx})^{-1} = 4.4$. This results in a composite fermion mean free path of $l_{cf} = g_{cf}/k_F = 46$ nm, where $k_F = \sqrt{2 \pi n}$. Thus, whilst the normal-electron properties infer that the Coulomb drag in our structures is ``ballistic'', with $l/d = 200$, the properties of composite fermions suggest that the drag will be much more diffusive in nature, with $l/d = 0.92$.

The calculation of the expected variance in Eq. \ref{eqVarDiffDrag} depends on the knowledge of the dephasing length of the composite fermions $L_\phi^{cf}$. However, the matter of the dephasing length in composite fermion systems is a non-trivial one. Dephasing occurs not only via \emph{e-e} interaction but also via interactions between electrons and the Chern-Simons gauge-field (see, e.g., \cite{LudwigPRB08} and references therein). In theory \cite{NarozhnyPRL01} phase breaking is assumed to be dominated by the latter mechanism: scattering of composite fermions by thermal fluctuations of the gauge field. The resulting dephasing length is $L_\phi \approx \sqrt{2 \pi e^2 d/ k_B T \epsilon}$, in contrast to the usual expression for the dephasing length of $L_\phi^{cf} = \alpha \sqrt{D\hbar g_{cf} / k_B T \ln{g_{cf}}}$, which comes from dephasing by \emph{e-e} scattering at low temperatures \cite{AltshulerJPhysC82}.

\begin{figure}[h]
\begin{center}\leavevmode
\includegraphics[width=1.0\linewidth]{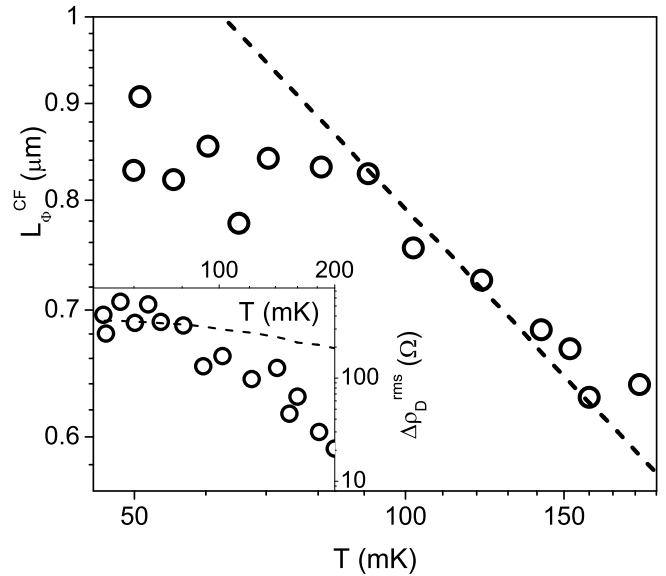}
\caption{$T$ dependence of $L_\phi$ found from the correlation concentration of the drag resistivity fluctuations. Dashed line is a plot of the calculated values of $L_\phi^{cf}$ assuming dephasing is dominated by \emph{e-e} scattering \cite{AltshulerJPhysC82} and using $\alpha = 1.7$. Inset: The amplitude of drag resistivity fluctuations plotted against $T$; $n = 1.45\times 10^{11}\,\mathrm{cm^{-2}}$. Dashed line is a theoretical plot of the amplitude using Eq. \ref{eqVarDiffDrag} and introducing a prefactor of 20.}
\label{Stats}
\end{center}
\end{figure}

The measured $L_\phi^{cf}$ (found from the correlation concentration $\Delta n_c = 1/2L_\phi^2$) is shown as a function of $T$ in Fig. \ref{Stats}. If one uses the expression for $L_\phi$ that comes from fluctuations of the gauge field then one obtains the correct temperature dependence, but the values are an order of magnitude too big compared with experiment. Our result is in agreement with the predictions of \cite{LudwigPRB08} on dephasing in composite fermion systems where, in the presence of long-range interactions, $L_\phi$ is expected to be well described by \emph{e-e} scattering in the limit of low temperatures, $T \ll 100/g_{cf}^2\tau_{cf}$, which for our system (with a low conductance of composite fermions) applies already below 35 K.

Using these values of $L_\phi^{cf}$ we calculate the expected variance of the drag resistivity fluctuations using Eq. \ref{eqVarDiffDrag}, which is plotted as the dashed line in the inset to Fig. \ref{Stats}, multiplied by 20 for the sake of comparison. We see that the magnitude of the fluctuations is underestimated when calculated using Eq. \ref{eqVarDiffDrag}, and the temperature dependence is poorly described at higher temperatures. However, this discrepancy between the experiment and the predictions of the diffusive drag theory in \cite{NarozhnyPRL01} is less than that previously seen for the case of Coulomb drag of normal electrons \cite{PriceScience07}, where fluctuations were four orders of magnitude larger in amplitude than that expected theoretically. This can be accounted for by our system being closer to the diffusive limit, $l/d < 1$.

\begin{figure}[h]
\begin{center}\leavevmode
\includegraphics[width=1.0\linewidth]{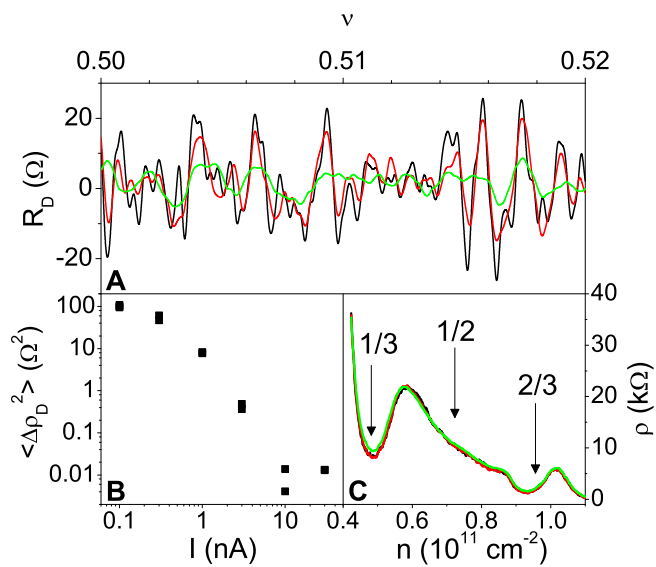}
\caption{Panel A: Drag resistivity as a function of filling factor measured with different active-layer currents: $I$ = 0.1 nA, 0.3 nA, and 1 nA. Panel B: variance of the drag resistance fluctuations as a function of driving current; $T$ = 50 mK, $n$ = 1.45$\times 10^{11}\,\mathrm{cm^{-2}}$. Panel C: single-layer resistivity as a function of filling factor measured at the same three currents, at $T$ = 50 mK, and $n$ = 0.57$\times 10^{11}\,\mathrm{cm^{-2}}$.}
\label{nonlinearity}
\end{center}
\end{figure}

The found value of $L_\phi$ is seen to deviate at low temperatures from that expected from \emph{e-e} scattering, Fig. \ref{Stats}. It is possible that this effect is related to a non-linearity of the drag fluctuations that we have observed at low temperatures. The fluctuations of the drag at $\nu =1/2$ are found to be strongly nonlinear, unlike in the case of weak magnetic fields, where both the average drag resistance and the fluctuations of the drag resistance were seen to be independent of the active-layer current \cite{PriceScience07}. The drag resistivity as a function of $\nu$ measured using different currents is shown in Fig. \ref{nonlinearity}A. The amplitude of the fluctuations increases by four times in decreasing the current from 1 nA to 0.1 nA. The nonlinearity of the fluctuations is stronger at higher currents, as demonstrated in Fig. \ref{nonlinearity}B, where the variance of the fluctuations is plotted as a function of current. This nonlinearity is not simply due to Joule heating, as the single-layer resistance is seen to be independent of driving current below 1 nA (Fig. \ref{nonlinearity}C). (Strong nonlinearity of the Coulomb drag of composite fermions was also seen in previous measurements of the average drag resistance \cite{LillyPRL98}.) The origin of this nonlinearity deserves further investigation in the future. All of the measurements we present in this paper were performed using a 0.1 nA driving current, where the nonlinearity is weak (Fig. \ref{nonlinearity}B).

To summarize, we have seen reproducible fluctuations of the Coulomb drag between composite fermions when varying carrier concentration in the two layers and magnetic field. There is a large enhancement in the size of fluctuations relative to that seen in drag between normal electrons, as was predicted theoretically. At low temperatures the magnitude of the fluctuations exceeds the average drag, such that the sign of the drag changes randomly with varying $n$ and $B$. The decoherence length found from the quasiperiod of the drag fluctuations is close to that described by \emph{e-e} scattering. The magnitude of the drag fluctuations is found to exceed that expected from the theory developed for the diffusive regime, though to a lesser extent than that seen in the case of drag fluctuations between normal electrons due to the shorter mean free path of composite fermions.

We would like to acknowledge the contributions to the fabrication of our samples by G. Allison, and to thank I. L. Aleiner, I. V. Gornyi, A. Kamenev, B. N. Narozhny, and Ady Stern for useful discussions.

\end{document}